\documentclass[twocolumn]{aastex631}

\shorttitle{}
\shortauthors{FU, HUANG, $\&$ YU}

\usepackage{amsmath} 
\usepackage{graphicx}
\usepackage{booktabs} 
\usepackage{tabularx}
\usepackage{mathrsfs} 
\usepackage{subfigure}
\usepackage[mathscr]{eucal}
\graphicspath{{./}{figures/}}

\begin{document}

\title{Boundary Layers of Circumplanetary Disks around Spinning Planets \\ I. Effects of Rossby Waves}

\correspondingauthor{Cong Yu}
 \email{yucong@mail.sysu.edu.cn}
\author[0000-0002-9108-9500]{Zhihao Fu}
\affiliation{School of Physics and Astronomy, Sun Yat-Sen University, Zhuhai 519082, China}
\affiliation{CSST Science Center for the Guangdong-Hong Kong-Macau Greater Bay Area, Zhuhai 519082, China}
\affiliation{State Key Laboratory of Lunar and Planetary Sciences, Macau University of Science and Technology, Macau, China}
\author[0000-0002-7276-3694]{Shunquan Huang}
\affiliation{School of Physics and Astronomy, Sun Yat-Sen University, Zhuhai 519082, China}
\affiliation{CSST Science Center for the Guangdong-Hong Kong-Macau Greater Bay Area, Zhuhai 519082, China}
\affiliation{State Key Laboratory of Lunar and Planetary Sciences, Macau University of Science and Technology, Macau, China}
\author[0000-0003-0454-7890]{Cong Yu}
\affiliation{School of Physics and Astronomy, Sun Yat-Sen University, Zhuhai 519082, China}
\affiliation{CSST Science Center for the Guangdong-Hong Kong-Macau Greater Bay Area, Zhuhai 519082, China}
\affiliation{State Key Laboratory of Lunar and Planetary Sciences, Macau University of Science and Technology, Macau, China}




\begin{abstract}
Gas giant planets are believed to accrete from their circumplanetary disks (CPDs). The CPDs usually involve accretion through the boundary layer (BL) in the vicinity of planets. Prior studies have concentrated on the BL of non-spinning planets. We investigate the influence of planetary spin on the wave behaviors within the BL. The rotation profile in such BLs would show sharp transition from the rigid rotation to the Keplerian rotation. We examine the angular momentum transport in these BL in terms of linear perturbation analysis. We find that the global inertia-acoustic mode associated with spinning planets would give rise to the inflow of angular momentum and the accretion of gas. In this work, we identify a new kind of global mode, namely the Rossby mode. The Rossby mode can lead to the outflow of angular momentum and the decretion of gas from a spinning planet. The Rossby mode provide a negative feedback that regulates the planetary spin and mass. We compare the growth rate of the two modes as a function of the width of BL, the Mach number and the spin rate of planet. Our results reveal the underlying hydrodynamic mechanism of terminal spins and asymptotic mass of the giant planets. 

\end{abstract}

\keywords{accretion, accretion disks  - planet formation - hydrodynamics - waves - instabilities}


\section{Introduction} \label{sec:intro}

The spin evolution is an important process during the accretion stage of many astrophysical objects, such as neutron stars, white dwarfs, protostars and protoplanets. If the accretion proceeds without any braking mechanism, the central bodies are thought to spin up to near-breakup velocities $\Omega_b=\sqrt{GM_{*}/R_{*}^{3}}$. However, $\Omega_* / \Omega_b$ for planets in our solar system can be only up to 0.4 (e.g. for Saturn). Further more, measurement of rotation rate by high-resolution spectroscopy indicates that some extrasolar planetary spins with the ration $\Omega_*/\Omega_b$ in the range of 0.08 – 0.3 \citep{2018NatAs...2..138B}. 
There is no significant correlation between the rotation rate and mass in the $1-20$ $M_{\rm Jup}$ range, and the spin rates are relatively stable during the initial epoch at the end of accretion. 
Thus, it is likely that there is a general physical mechanism involving interactions between protoplanets and CPDs, which determines the terminal spins during the accretion phase.

For stars in circumstellar disks, the high temperature strongly ionizes materials at the inner disk, making them highly coupled to the magnetic fields. The low spin rate of central bodies can be attributed to magnetic braking \citep{1996MNRAS.280..458A} or stellar wind \citep{2005ApJ...632L.135M}. Although \cite{2009Natur.457..167C} argued that proto-Jupiters can generate much stronger fields, current observations suggest that Jupiter's magnetic field is only about 4 G 
\citep{1974JGR....79.3501S}, which is not strong enough to support the above mechanisms. 

In cases of weakly magnetized or pure hydrodynamic models, accretion from circumplanetary disks to planets will inevitably experience a sharp decrease in rotation profile. Such a transitional region from Keplerian rotation to rigid rotation is known as
$\emph{boundary layer}$ (BL). Since the angular momentum of materials in a Keplerian disk are proportional to $r^{1/2}$, they must lose their angular momentum in order to fall into planets. While the loss of angular momentum has been solved by the magnetorotational instability (MRI, \citealt{1991ApJ...376..214B}), it cannot be applied into the BL. This is because $d\Omega/dr > 0 $ within the BL is in contradiction with the requirement $d\Omega / d r < 0 $ of MRI, and it has been verified by \cite{2013ApJ...770...68B}.

Measurement of transits at multiple wavelengths has uncovered the existence of atmosphere on exoplanets such as hot Jupiters \citep{2015AREPS..43..509H}. The planetary atmosphere also plays an essential role in the dynamical evolution of planets. 
Various waves can propagate in the atmosphere, such as acoustic-gravity waves or inertia-gravity waves \citep{vallis_2017}. Perturbations can also propagate in CPDs in the form of density waves \citep{1978ApJ...222..850G}.
Considering the planet and the disk as a whole, these local waves form a global mode with a pattern speed $\Omega_P$ that cannot be characterized by local effective viscosity \citep{1973A&A....24..337S}.
When the central object is non-rotating, \cite{2012ApJ...752..115B} identified three types of acoustic modes named as the upper, the middle and the lower mode \citep{2013ApJ...770...67B}.  Acoustic waves are excited within the BL and then propagate into the planet and the disk, giving rise to the angular momentum transport on a global scale.
\cite{2022MNRAS.509..440C,2022MNRAS.512.2945C} studied the spatial morphology of these waves in detail and discovered some new vortex-driven modes in the vicinity of the BL. 
They also found that the formation of vortices and the formation of spiral arms can be explained by Rossby wave instability (RWI) \citep{1999ApJ...513..805L,2000ApJ...533.1023L}. 

When it comes to a spinning planet, there are two different ways to explain why the BL can limit the spin of a central object. \cite{1993PhDT........41P} supposed that the BL itself (defined by the existence of a maximum in angular velocity) ceases to exist, so that a conventional viscous stress can transport angular momentum outward from the central object into the disk. \cite{2021ApJ...921...54D} confirmed the effect by carrying out viscous hydrodynamic simulations in 2D ($\emph{r}$ - $\theta$) axisymmetric geometry, who also restricted the terminal spins to  $\Omega_{*}/ \Omega_{b} \sim 0.6-0.8$.
On the other hand, the angular momentum transport in the boundary layer may be fundamentally non-viscous
(e.g. via waves), thereby outward transport from the planet can coexist with an angular velocity profile that has the BL. 
In this regime, \cite{2021MNRAS.508.1842D} found that angular momentum transport is inefficient at high spin rate, and the value decreases by $\sim 1-3$ orders of magnitude at $\Omega_{*} -0.4 \sim 0.6$. 
While numerous numerical simulations show that waves are indeed important in constraining the upper limit of spin rate, the underlying origin is still unclear and more detailed analysis are needed. In this work, we propose a physical mechanism to explain the phenomena in the BL of spinning planets within CPDs.

This paper is organized as follows. In $\S$\ref{Basic} we display the perturbation equations and derive the dispersion relations for planets and disks. In $\S$\ref{setup} we display our physical setup and solving methods. In $\S$\ref{Global modes} we show our results and elucidate the global mode in detail. In $\S$\ref{Discussion} we discuss the astrophysical implications of our results and how they are correlated to the numerical simulations. In $\S$\ref{Summary} we summarize the main conclusions.

\section{Basic Theory}\label{Basic}

\subsection{Fundamental Equations}

The nonlinear hydrodynamic equations can be written as follows 
\begin{equation}
\frac{\partial \rho}{\partial t}+\nabla \cdot (\rho \textbf{v})=0 \ , 
\end{equation}

\begin{equation}
\frac{\partial \textbf{v}}{\partial t}+
\textbf{v}\cdot \nabla \textbf{v}
=-\frac{1}{\rho}\nabla P
-\nabla \Phi \ , 
\end{equation}

\begin{equation}\label{isentropy}
\frac{\partial P}{\partial t}+
\textbf{v}\cdot\nabla P+
\gamma P \nabla \cdot \textbf{v}=0 \ .
\end{equation}
They correspond to the mass conservation equation, the momentum equation and adiabatic energy equation, respectively. Equation \eqref{isentropy} is equivalent to $\frac{D}{D t} (P/\rho^{\gamma})=0$.

\subsection{Linear Perturbation Analysis}
We adopt the cylindrical coordinates  $(r,\phi,z)$ for circumplanetary disks. The equilibrium state of the CPD can be written as 
\begin{equation}\label{equilibrum}
\frac{V^2}{r} \equiv
\Omega^{2}r
=\frac{1}{\rho}\frac{\partial P}{\partial r}
+\nabla \Phi \ .
\end{equation}
The gravitational potential is dominated by the central planet, which is
\begin{equation}
    \Phi=-\frac{1}{r} \label{potential} \ .
\end{equation}
For simplicity, we do not consider the self-gravity of CPDs. 
All perturbations are denoted with a prefix $\delta$. For arbitrary physical variables, $f$, the perturbations can be Fourier decomposed as $\delta f \propto f(r)\exp [i(m \phi + k_z z- \omega t)]$. We get the linearized equations
\begin{equation}\label{bl1}
i\sigma\delta\rho=
\frac{1}{r}\frac{\partial (\rho r)}{\partial r}
\delta v_{r}+
\rho\frac{\partial \delta v_{r}}{\partial r}+
i k_{\phi}\rho \delta v_{\phi}+
i k_{z}\rho \delta v_{z} \ ,
\end{equation}
\begin{equation}
-i\sigma \delta v_{r}-2\Omega\delta v_{\phi}=
-\frac{1}{\rho}\frac{\partial \delta P}{\partial r}+
\frac{\delta\rho}{\rho^{2}}\frac{\partial P}{\partial r} \ ,
\end{equation}
\begin{equation}
-i\sigma \delta v_{\phi}+\frac{\kappa^{2}}{2\Omega}\delta v_{r}=
-i\frac{k_{\phi}}{\rho}\delta P \ ,
\end{equation}
\begin{equation}
-i\sigma \delta v_{z}=
-ik_{z}\frac{\delta P}{\rho} \ ,
\end{equation}
\begin{equation}\label{bl5}
i\sigma \delta P=i\sigma c_{s}^{2}\delta \rho+
\frac{\rho c_{s}^{2}}{L_{S}}\delta v_{r} \ ,
\end{equation}
where $\sigma = \omega - m \Omega $ is the local Doppler-shifted frequency, $k_{\phi}=m / r$ is the azimuthal wave vector ,
$c_{s}^{2}=\gamma P / \rho$ is the sound velocity, 
\begin{equation}\label{epicyclic}
    \kappa^2\equiv \frac{1}{r^3}\frac{d(\Omega^2 r^4)}{d r}=
    4 \Omega^2 + 2 \Omega \frac{d \Omega}{d r}r \ ,  
\end{equation}
is the square of radial epicyclic frequency. Additionally, we have introduced the length scale of entropy variation $L_S$ which satisfies
\begin{equation}
    \frac{1}{L_S}= \frac{1}{\gamma}\frac{d}{dr} \left[\ln \left( \frac{P}{\rho^{\gamma}} \right) \right] \ .
\end{equation}
Note that for the adiabatic process, the spatial scale $L_{ S} \rightarrow \infty$.   
Similarly, for the length scale of pressure and density can be written as
\begin{equation}
    \frac{1}{L_{P}}=\frac{1}{\gamma P}  \frac{d P}{d r}  \ ,  \
    \frac{1}{L_{\rho}}=\frac{1}{\rho}  \frac{d \rho }{d r} \ ,
\end{equation}
respectively. Note that $(L_S)^{-1}=(L_P)^{-1}-(L_{\rho})^{-1}$ \citep{1999ApJ...513..805L}.
If we define $\Psi=\delta P/\rho$, then the above equations can be cast as two first order linear equations
\begin{equation}\label{linear1}
\frac{\partial \delta v_{r}}{\partial r}+
A_{11}\delta v_{r}+A_{12}\Psi=0 \ ,
\end{equation}
\begin{equation}\label{linear2}
\frac{\partial \Psi}{\partial r}+
A_{21}\delta v_{r}+A_{22}\Psi=0 \ ,
\end{equation}
where
\begin{equation}
A_{11}=\frac{1}{L_{\rho}}+\frac{1}{L_{S}}+
\frac{1}{r}+\frac{k_{\phi} \kappa^{2}}{2\Omega \sigma}  \ , 
\end{equation}
\begin{equation}
A_{12}=i\left(\frac{k_{\phi}^{2}}{\sigma}-\frac{\sigma}{c_{s}^{2}}+
\frac{k_{z}^{2}}{\sigma} \right)   \ ,
\end{equation}
\begin{equation}
A_{21}=-i\left(\sigma-\frac{\kappa^{2}}{\sigma}+\frac{c_s^2}{\sigma L_{S} L_{P}} \right)    \ ,
\end{equation}
\begin{equation}
A_{22}=-\frac{1}{L_{S}}-\frac{2\Omega k_{\phi}}{\sigma} \ .    
\end{equation}
They are equivalent to the equations derived in \citet {2012ApJ...752..115B}.

\subsection{Dispersion Relations}\label{Dispersion_relation}
In this paper, we shall focus on the 2D perturbations with $k_z=0$.  Three dimensional effects will be investigated further and reported elsewhere. 
\subsubsection{Waves in Rotating Stratified Planetary Atmospheres}
 
 When we study the properties of waves, it is useful to adopt a local reference frame which transforms the cylindrical coordinate $(r,\theta,z)$ into the Cartesian coordinate $(x,y,z)$. Therefore, we can ignore the curvature term and have the same form as it is in \cite{2017ApJ...835..238B}
\begin{equation}\label{Cardp}
    i \left(\frac{\sigma^2}{c_s^2} - k_y^2 \right) \Psi =
    \sigma \delta v_x ' +
    \sigma \left(\frac{1}{L_S}+\frac{1}{L_{\rho}}+ \frac{k_y \kappa^2}{2\Omega \sigma} \right) \delta v_x \ , 
\end{equation}
\begin{equation}\label{Carvx}
    i \left(\sigma^2-\kappa^2+\frac{c_s^2}{L_S L_p} \right) \delta v_x =
    \sigma \Psi'-\sigma \left(\frac{1}{L_S}+\frac{2 \Omega k_y}{\sigma} \right) \Psi  \ .
\end{equation}
where the prime $^{\prime}$ denotes the derivative with respect to $x$.
 
 The hydrostatic equilibrium state of the planet atmosphere satisfies
 \begin{equation}\label{stratification}
     \frac{d P}{d x}= - \rho g  \  ,  \ 
     \frac{\rho(x)}{\rho_0}=\frac{P(x)}{P_0}=e^{-{x}/{H}} \ ,
 \end{equation}
where $H = RT/g = c_s^2/(\gamma g)$ is the atmosphere scale height, $T$ is the temperature of atmosphere and $R$ is the universal gas constant. Here $g=\nabla \Phi - \Omega^2 r$ characterizes the gravitational acceleration revised by the centrifugal force. Substituting it into \eqref{Cardp} and \eqref{Carvx}, we obtain 
\begin{equation}\label{atm1}
    i \left( \frac{\sigma^2}{c_s^2} - k_y^2 \right) \Psi =
    \sigma v_x' +
    \sigma \left( - \frac{g}{c_s^2}+ \frac{k_y \kappa^2}{2 \Omega \sigma} \right) \delta v_x  \  ,
\end{equation}
\begin{equation}\label{atm2}
    i \left(\sigma^2 - \kappa^2 - N^2\right) \delta v_x =
    \sigma \Psi' - \sigma \left(\frac{N^2}{g} + \frac{2\Omega k_y}{\sigma} \right) \Psi \ ,
\end{equation}
where $N^2=(\gamma-1)g^2/c_s^2$ is the Brunt-Vaisala frequency. Following \cite{1960CaJPh..38.1441H}, we put
\begin{equation}\label{evanecent_wave}
    \Psi \ , \  \delta v_x \sim 
    \exp\left(\frac{x}{2H}+i k_x x \right) \ .
\end{equation}
Besides, for a rigid body rotation ($\Omega =$ const), we have $\kappa^2=4 \Omega^2$ according to equation \eqref{epicyclic}. Combining \eqref{atm1} and \eqref{atm2} and then eliminating $\Psi$ and $\delta v_x$, we can obtain the following dispersion relation\footnote{It is interesting to note that when rotation is ignored, acoustic modes and internal gravity modes can be readily recovered in this equation.}  
\begin{equation}
\begin{split}
    \sigma^4 + C_2\sigma^2 + C_1 \sigma  + C_0 =0 \ .
\end{split}
\end{equation}
where 
\[
C_2 = - \left[ c_s^2\left(k_x^2+k_y^2 \right) + \frac{c_s^2}{4 H^2}+ 4\Omega^2 \right] \ , 
\]
\[
C_1 = 2(2-\gamma)g \Omega k_y \ , 
\]
\[
C_0 = k_y^2 g^2 (\gamma-1) = k_y^2 N^2 c_s^2 \ .
\]
For the isothermal condition $(\gamma =1)$, the above equation reduces to
\begin{equation}\label{inner dispersion}
\begin{split}
    \sigma^3 + C_2 \sigma + 2g \Omega k_y = 0 \ .
\end{split}
\end{equation}
The high frequency branch of the above equation is the inertia-acoustic mode, whose dispersion relation can be written as 
\begin{equation}
\sigma^2 = \left[ c_s^2\left(k_x^2+k_y^2 \right) + \frac{c_s^2}{4 H^2}+ 4\Omega^2 \right] \  , 
\end{equation}

It is clear from this equation that the inertia-acoustic modes have been modified by the gravity and rotation. Note that for a non-spinning planet $(\Omega = 0)$,  this dispersion relation recovers the result in \citet{2013ApJ...770...68B}. 
In this paper we identify a new regime for low frequency wave propagation in BLs of CPDs. The low frequency branch dispersion relation can be written as
\begin{equation}
    \sigma = \frac{ (2/\gamma) H \Omega k_y }{H^2\left(k_x^2+k_y^2 \right) + \frac{1}{4}+ (4\Omega^2 H^2/c_s^2) }  \  . 
\end{equation}
We note that this dispersion relation is pretty similar to the Rossby wave, namely
\begin{equation}
    \omega = \frac{-\beta k}{K^2},
\end{equation}
where $\beta = {df}/{dy}$ (Coriolis parameter f) and $K$ is the total wavenumber.
The difference comes from the fact that we consider the effects of stratification and rotation, and we call this new branch as Rossby mode throughout this paper.

\subsubsection{Waves in Disks}
Provided that the density is uniform $(\rho= \rm const)$, the equations \eqref{Cardp} and \eqref{Carvx} can be simplified as
\begin{equation}\label{disk1}
    i \left(\frac{\sigma^2}{c_s^2}-k_y^2 \right)\Psi =
    \sigma \delta v_x' +
     \frac{k_y \kappa^2}{2\Omega}\delta v_x  \ , 
\end{equation}
\begin{equation}\label{disk2}
    i \left(\sigma^2-\kappa^2 \right)\delta v_x=
    \sigma  \Psi' -
    2 \Omega k_y \Psi  \  .
\end{equation}
With the WKB approximation, $\Psi$, $\delta v_x \sim \exp (i k_x x)$,
and then combining \eqref{disk1} and \eqref{disk2}, we obtain the dispersion relation
\begin{equation}\label{disk_relation}
    \sigma^3-\left[c_s^2(k_x^2+k_y^2)+
    \kappa^2 \right]\sigma+
    i\left(\frac{\kappa^2}{2\Omega}-2\Omega \right)k_x k_y c_s^2 =0  \ .
\end{equation}
The rotation profile of the disk is Keplerian $(\Omega \sim r^{-3/2})$ due to the central potential field, so we have $\kappa^2 = \Omega ^2$ according to equation \eqref{epicyclic}. In the tight-winding limit $(k_x \gg k_y)$, we can ignore the term involving $k_y$ and using
$\sigma=\omega-m\Omega$ to arrive at
\begin{equation}\label{Outer dispersion}
(\omega- m\Omega)^2
= c_s^2k_x^2+\kappa^2 \ .
\end{equation}
This defines the dispersion relation for spiral density waves \citep{1978ApJ...222..850G}.


\subsection{Group Velocity}
By the definition of group velocity, we know that
\begin{equation}
    \textbf{c}_{g}=\frac{\partial \omega}{\partial \textbf{k}}=
    \frac{\partial \sigma}{\partial \textbf{k}} \ .
\end{equation}
For the inertia-acoustic modes, the group velocity along $x$ direction in the planetary atmosphere reads
\begin{equation}\label{Group_velocitx}
    c_g^x = \frac{ k_x c_s^2 }{ \left[c_s^2(k_x^2+k_y^2)+\frac{c_s^2}{4H^2} + 4 \Omega^2 \right]^{1/2}} \ . 
\end{equation}
It is clear that the group velocity along the $x$ direction is the same as the sign of $k_x$ for the inertia-acoustic modes. 
For the Rossby modes, the group velocity in the $x$ direction in the planetary atmosphere reads 
\begin{equation}\label{Rossby cgx}
    c_g^x = - \frac{(4 c_s^2 g \Omega k_y) k_x }{\left[c_s^2(k_x^2+k_y^2)+\frac{c_s^2}{4H^2} + 4 \Omega^2 \right]^2} \ .
\end{equation}
We can readily know that the group velocity along the $x$ direction is the opposite to the sign of $k_x$ for the Rossby modes.
From equation \eqref{Outer dispersion}, we know that the group velocity along the $x$ direction in the disk reads 
\begin{equation}\label{Group_velocitx2}
    c_g^x = \frac{k_x c_s^2}{\omega - m \Omega} \ .
\end{equation}
For the density wave to flow outside the disk, the waves need to be trailing, i.e., $k_x$ $>$ 0. 

\section{Physical setup and solving method}\label{setup}

\subsection{Mach Number}
Assuming that the radius of planet $r_*=1$, and the angular velocity $\Omega(r_*) = 1$. Therefore, we can define the characteristic Mach number as 
\begin{equation}
     \mathcal{M} \equiv V(r_*)/c_s = c_s^{-1}.
\end{equation}
From equation \eqref{stratification} we can know that the scale height of isothermal atmosphere
$
    H = c_s^2/ g(r)
$
, which means the thickness of atmosphere will decrease sharply with the increase of the Mach number.
In addition, \cite{1998RvMP...70....1B} has given that for steady-state Keplerian disks, the scale height of the disk in the vertical direction satisfies $h= \sqrt{2}c_s/\Omega (r)$, so
$
    H = h(r_*) \mathcal{M}^{-1}/2 \ll h(r_*).
$

\subsection{Initial Rotation and Density Profiles}

\begin{figure}[t]
    \centering
    \includegraphics[width=1\linewidth]{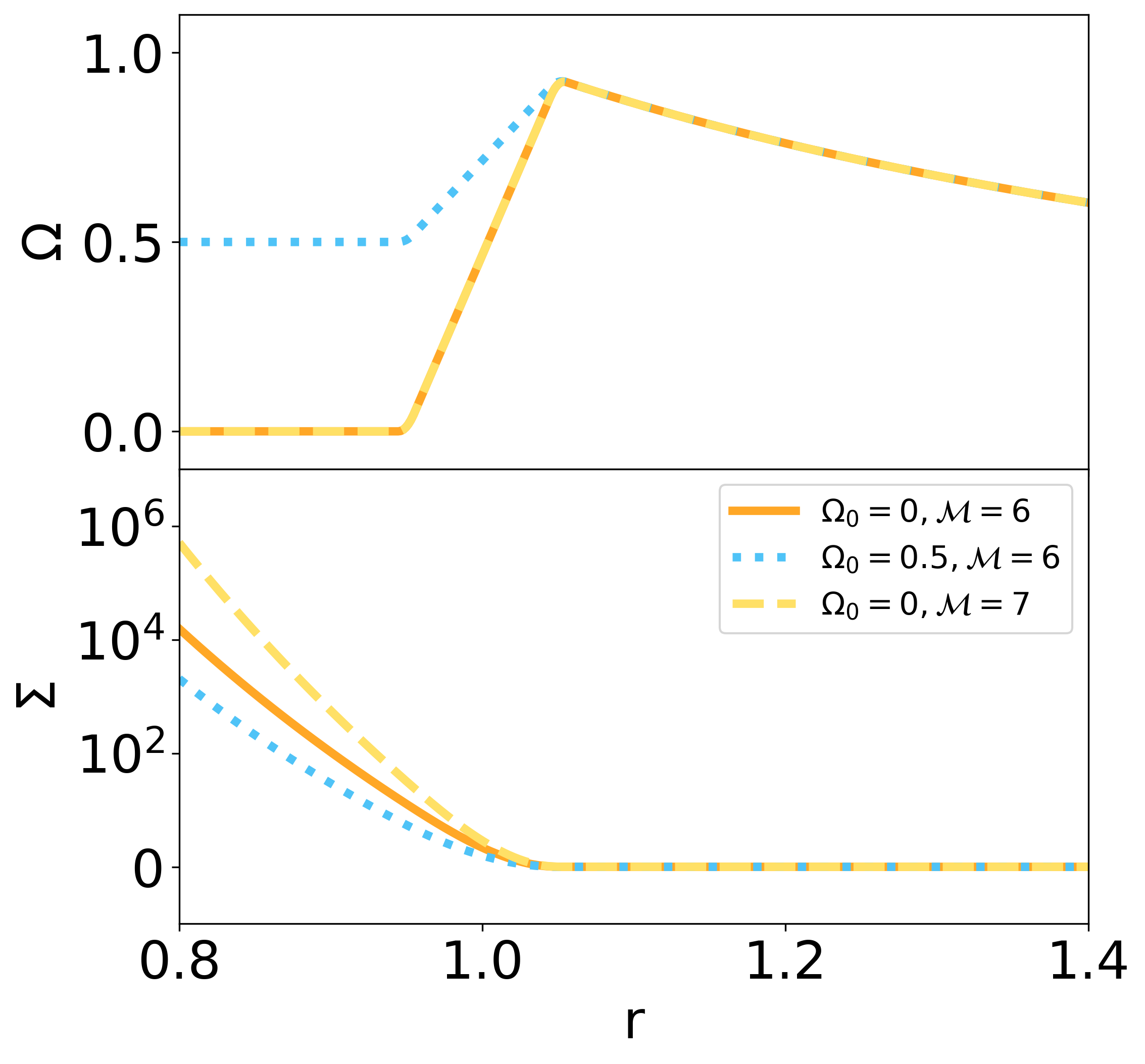}
    \caption{Equilibrium state for three initial conditions corresponding to $\delta r=0.05$, where $\Omega_0$ is the rotation velocity of planet and $\mathcal{M}$ is the Mach number.}
    \label{surface_density}
\end{figure} 

As we have assumed in section \eqref{Dispersion_relation}, the disk follows the Keplerian rotation profile while the planet maintain rigid body rotation. Within the boundary layer, we use a linear function that transitions smoothly at the junctions. Hence, the rotation curve can be described as  
\begin{small}
\begin{equation}
\Omega(r)=\left \{
\begin{array}{lcl}
\Omega_0 & & 
{r < r_* - \delta r}\\
r^{-3/2}& & 
{r >r_* + \delta r}\\
\Omega_0 + \left[(r_*+ \delta r)^{-3/2}-\Omega_0 \right]
\frac{r+\delta r - r_*}{2 \delta r}& & 
{r_* - \delta r \leq r}\\
&&  {\leq r_* + \delta r} \ ,
\end{array} \right.
\end{equation}
\end{small}
where $\delta r$ is the width of boundary layer and $\Omega_0$ is the spin rate of planet. In addition, we assume that the surface density  $\Sigma$ maintains a fixed value throughout the disk. Thus, we can integrate the equilibrium state equation \eqref{equilibrum} to acquire the surface density within planetary atmosphere. Mathematically, it can be inferred that $\Sigma \propto \exp (\mathcal{M}^2/r)$. As shown in Figure \ref{surface_density}, it decays exponentially as the distance increases. A higher spin rate gives rise to a decrease in slope while a higher Mach number makes it steeper.

\subsection{Axisymmetric Stability}
To focus on the non-axisymmetric behavior of the boundary layers, it would be helpful to maintain the stability of equilibrium state disk under axisymmetric perturbation $(m=0)$. Due to the radial stratification and the rotation, the sufficient condition for local stability has been given by \cite{1978ApJ...220..279E}, which is also known as the Solberg-Hoiland criterion,
\begin{equation}
    \kappa^2(r)+N^2(r) \geq 0 \ .
\end{equation}
Since $N^2=0$ under the isothermal condition $\gamma=1$, it becomes the Rayleigh criterion 
\begin{equation}
    \kappa^2(r) \geq 0 \ .
\end{equation}
As the angular velocity distribution we have previously assumed, this condition is always satisfied. Hence, the equilibrium state is stable to the axisymmetric perturbation. 

\subsection{Methods of Solving Linear Eigenvalue Equations}
Noting that the coefficients of the first-order differential equations \eqref{linear1} and \eqref{linear2} are complex, which are also the functions of $r$. Hence, the eigenfrequency $\omega$ are solutions on the complex plane. In this paper, we adopt a relaxation method to solve the problem \citep{1992nrfa.book.....P}. Briefly, this method can be separated into the following steps. Firstly, replacing the ODEs by approximate finite-difference equations on a grid or mesh of points that covers the region of interest. Then, giving it a initial guess solution and improving the result by Newton-Raphson technique. Meanwhile, iterations are carried out by a particularly designed Gaussian elimination. Finally, we can acquire the converged eigenvalue after enough iterations.

\begin{deluxetable}{cccccccccc}
\tablenum{1}
\tablecaption{Initial parameters and results \label{Results}}
\tablewidth{0pt}
\tablehead{
\colhead{m} & \colhead{$\delta r$}
& \colhead{$\mathcal{M}$} & \colhead{$\Omega_0$} &\colhead{$\Omega_P$} &\colhead{Im[$\omega$]} &\colhead{$k_{1}$}  & \colhead{$k_{2}$} 
&\colhead{$c_g^x$} 
}
\startdata
7 & 0.05 & 6 & 0 & 0.768 & 0.075 & -14.5 & 21.4 & -0.076  \\
15 & 0.05 & 6 & 0.5 & 0.830 & 0.179 & -10.4 & 51.5 & -0.067 \\
7 & 0.05 & 6 & 0.7 & 0.729 & 0.122 & -37.3 & 19.7 & 0.008\\
15 & 0.05 & 6 & 0.5 & 0.522 & 0.198 & -36.0 & 23.4 & 0.011\\
\enddata
\tablecomments{The columns from left to right are: mode number, width of BL, Mach number, spin rate of planet, pattern speed, growth rate, wavenumber at $r_1$, wavenumber at $r_2$, group velocity in the x direction at $r_1$.}
\end{deluxetable}
\subsection{Boundary Conditions}
In our model, we set $r_{1}=0.8$ and $r_{2}=2.5 $ as the inner and outer boundary, respectively.
The boundary conditions depend on the sign of $k_r$.
At the inner boundary, the wave mode can be inertia-acoustic mode or Rossby mode,
depending on whether the wave frequency lies in the high or low frequency range. At the outer boundary, the wave mode is always the density wave of the disk. We choose the appropriate $k_r$ according to the group velocity at the boundaries. 
In other words, we need ascertain the relation between the phase velocity and the group velocity based on the dispersion relations, and then choose the correct wave number $k_r$.

\subsection{Numerical Results}

The main parameters in BLs are the azimuthal wave number $m$, the width of BL $\delta r$, the Mach number $\mathcal{M}$ and the spin rate of planet $\Omega_0$. 
We have tried calculations with different resolutions and good numerical convergence is achieved in our solutions. 
Given an initial guess solution, a correct eigenvalue $\omega = \omega_r + i \omega_i$ can be achieved by numerical iterations. 
 Table \ref{Results} lists the initial parameters and corresponding results including the pattern speed $\Omega_P$, the growth rate Im[$\omega$] (imaginary part of the eigenvalue), and the radial wave number $k_{1}$ and $k_{2}$ for the inner and the outer boundary. The group velocities $v_{g}^{x}$ at the inner boundary are directly solved from the equation \eqref{inner dispersion}.

\section{Global modes of Boundary Layers}\label{Global modes}
We will discuss in further details about two global modes, i.e., inertia-acoustic and Rossby modes, for the spinning planets and study their spatial structure and response to various physical parameters.

\subsection{Inertia-acoustic and Rossby Modes}
 According to \citet{vallis_2017}, acoustic-gravity waves (p and g modes) exist in a stratified, compressible planetary atmosphere. The isothermal equation of state ($\gamma =1$) means $N^2 =0$, so g-modes are filtered out.
 An important property of acoustic waves is that the radial component of phase velocity is in the same direction as that of the group velocity, as shown in the first two rows of Table \ref{Results}.  
 From the dispersion relation of the planet, we know that there are two inertia-acoustic modes with opposite directions of propagation. Our calculations show that only those results with $\sigma > 0$ have physical meaning, so we ignore modes with $\sigma <0$. 
 In fact, this kind of acoustic global mode has been found and classified into three branches, namely upper, middle and lower \citep{2013ApJ...770...67B}. 
 The upper mode and the middle mode emerge during the linear phase, but the latter one with negative growth rate will decay rapidly. 
 The lower mode is evolved from the upper mode because of the nonlinear interactions between waves. 
 Therefore, our inertia-acoustic modes here correspond to the upper mode during the linear phase for non-rotating planets.
 
 When it comes to the low frequency regime, the new Rossby wave would come into play. 
Looking at the last the last two rows of Table \ref{Results}, the group velocities is opposite to the phase velocities in the $x$ direction,
The deviation between $\Omega_0$ and $\Omega_P$ is very small, so they are actually low frequency waves in a co-moving frame of planet. 
These features are consistent with the behaviours of Rossby waves, and we call them Rossby modes.

In the following parts, we will study these two modes in greater details. The results of the inertia-acoustic mode correspond to parameters $m=7$, $\delta r=0.05$, $\mathcal{M}=6$ and $\Omega_0=0$, and the results of the Rossby mode correspond to parameters $m=7$, $\delta r=0.05$, $\mathcal{M}=6$ and $\Omega_0=0.7$.
\begin{figure}[t]
    \centering
    \includegraphics[width=1\linewidth]{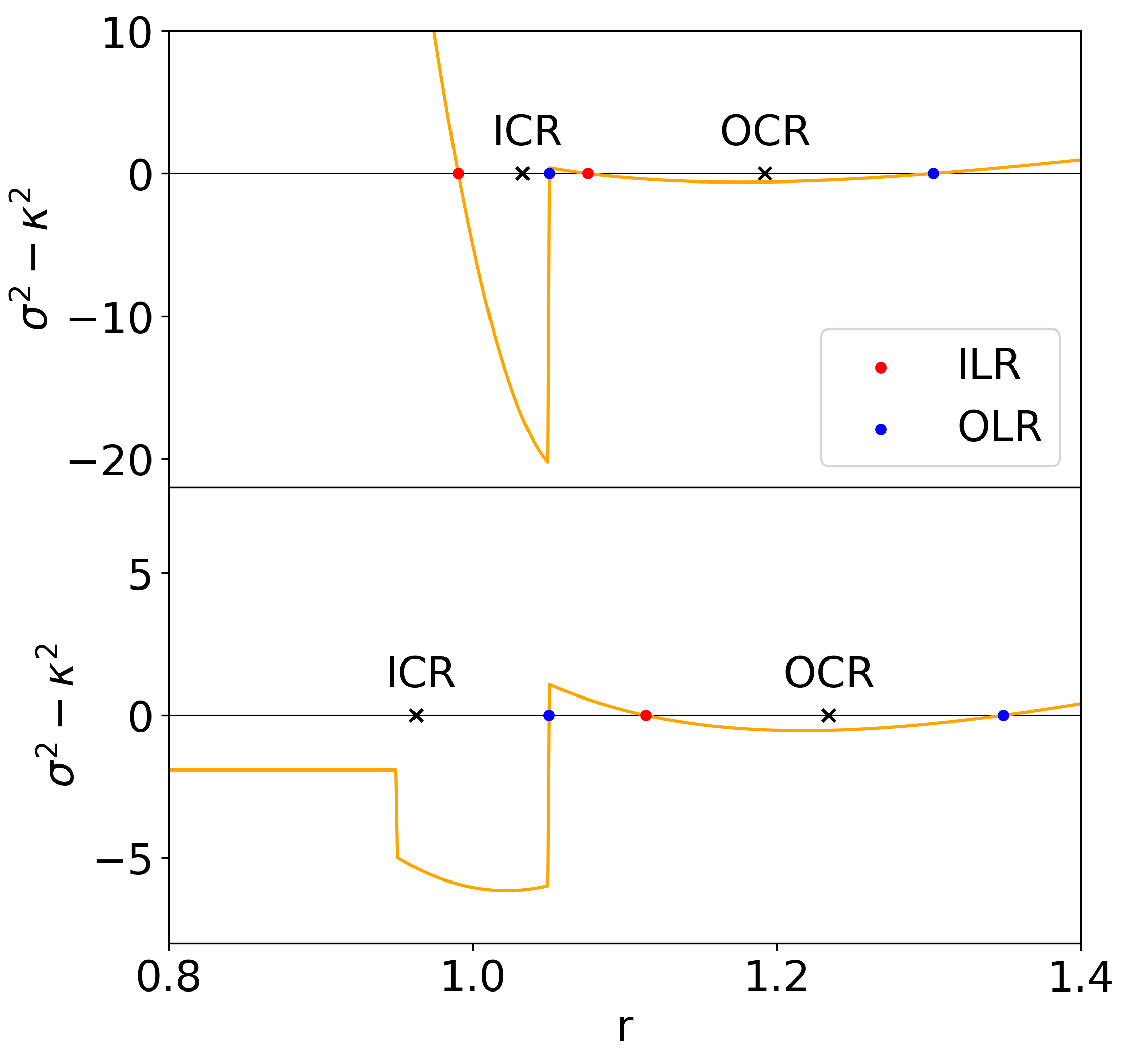}
    \caption{Diagrams of $\sigma^2-\kappa^2$ as a function of r. The upper panel corresponds to inertia-acoustic modes and the lower panel corresponds to Rossby modes. The cross indicates the corotation resonance (CR), the red circle indicates the inner Lindblad resonance(ILR) and the blue circle indicates the outer LindBlad resonance (OLR).   }
    \label{positions}
\end{figure} 
\begin{figure*}[t]
\begin{minipage}{0.45\textwidth}
    \centering
    \subfigure[]
    {
    \centering
    \includegraphics[width=1\linewidth]{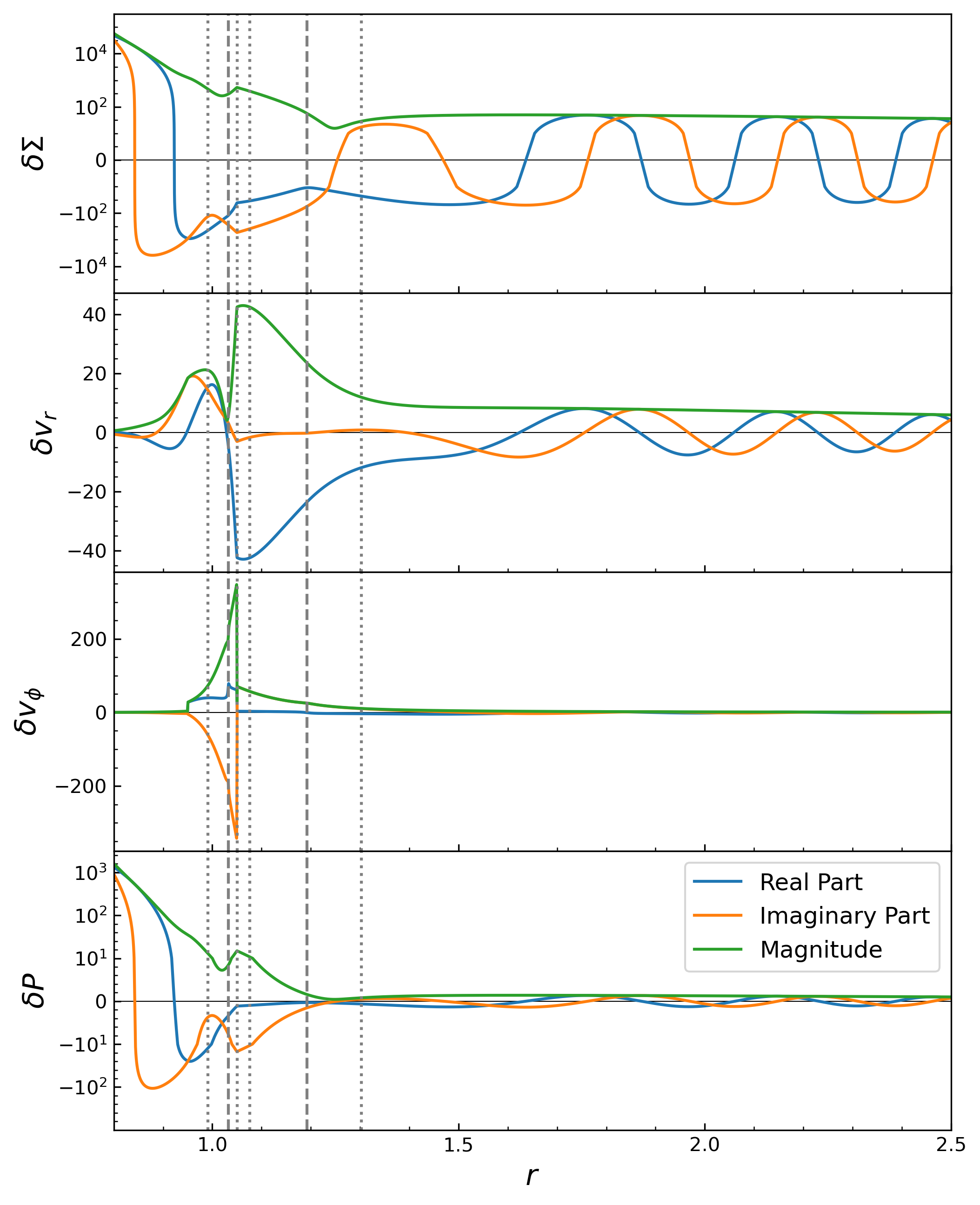}
    }
\end{minipage}
\begin{minipage}{0.45\textwidth}
     \subfigure[]
    {    
    \centering
    \includegraphics[width=1\linewidth]{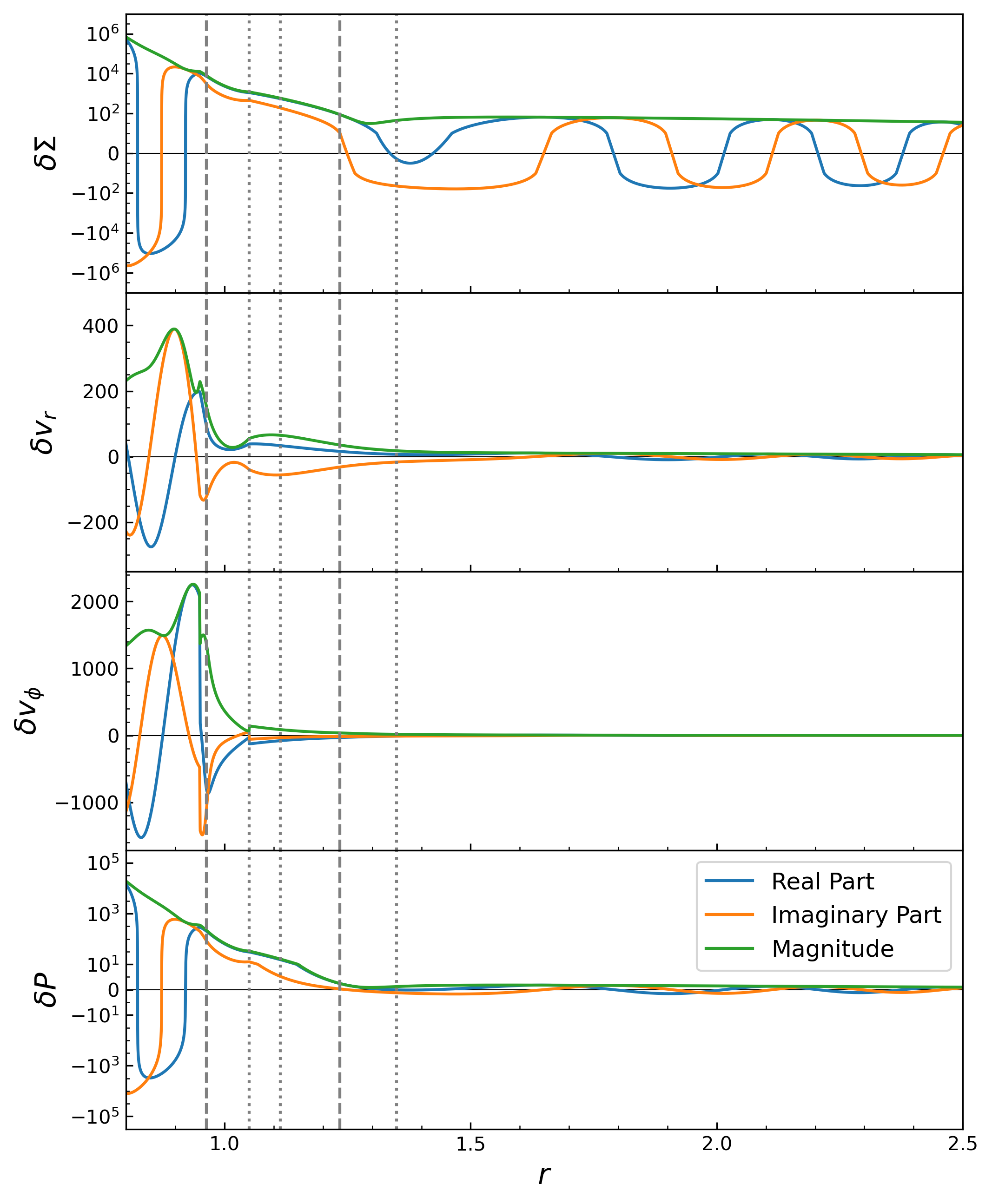}
    }
\end{minipage}
    \caption{Profiles of perturbations for inertia-acoustic modes in panel (a) and Rossby modes in panel (b). The blue lines are real part, the orange lines are imaginary part and the green lines are magnitude part. The dashed lines indicate the corotation resonance, and the dotted lines indicate the Lindblad Resonance.}
    \label{perturbation-profile}
\end{figure*} 
\subsection{Corotation and Lindblad Resonance}\label{CR and LR}
The corotation resonance (CR) is where the rotation speed is equal to the pattern speed $\Omega = \Omega_P$. We can see that there are always two corotation resonances if $\Omega_P>\Omega_0$ from Figure \ref{positions}. The inner corotation resonance (ICR) is inside the BL with $r \approx 1$, and the outer corotation resonance (OCR) is located at 
\begin{equation}
    r_{OCR} = \Omega_P^{-2/3} \ .
\end{equation}
The Lindblad resonances (LRs) appear where $\sigma^2=\kappa^2$. Their locations are determined by the following equation,
\begin{equation}\label{LR}
    r_{LR}=\left ( \frac{m}{m \pm 1} \Omega_P \right )^{-2/3} \ ,
\end{equation}
 From equation \eqref{Outer dispersion}, we know that
\begin{equation}
    k_r = \frac{1}{c_s} \sqrt { m^2 [\Omega - \Omega_{P}]^{2} - \kappa^2}.
\end{equation}
Note that $k_r$ becomes imaginary between of inner LR and outer LR, and waves could not propagate in these areas which are also referred as to the evanescent region. 
However, equation \eqref{LR} can only be applied into Keplerian disks, and it is invalid for rigid body rotation and the BL.
The other resonances can be found by $\sigma^2=\kappa^2$, i.e. zero points of $\sigma^2-\kappa^2$, as shown in Figure \ref{positions}. The inner LR are indicated in red circles and the outer LR are indicated in blue circles. 
Interestingly, each CR has a pair of LRs for inertia-acoustic modes
in the upper panel, whereas there is only one outer LR near the ICR for Rossby modes in the lower panel.
In the following, we will see how the difference affects the behaviour of two modes.

\subsection{Perturbation Quantities and Spatial Morphology}

The panel (a) and (b) in Figure \ref{perturbation-profile} show radial profiles of perturbations of inertia-acoustic modes and Rossby modes, respectively. We show the perturbations of the density $\delta \Sigma$, the radial velocity $\delta v_r$, the azimuthal velocity $\delta v_{\phi}$, and the pressure $\delta P$ from top to bottom. 
In this figure, blue lines indicate the real part, orange lines for the imaginary part and green lines for the magnitude. 

On the left panel, we can see that the magnitude of velocity perturbations $\delta v_r$ and $\delta v_\phi$ reach a maximum between the ICR and the OCR, with larger value in the disk. 
In contrast, on the right, the magnitude of velocity perturbations is larger in the planet.
According to our analysis in $\S$\ref{CR and LR}, there are evanescent regions near both the ICR and the OCR. Due to the super-reflection of the CR, majority of waves are trapped in such a resonant cavity formed by two CRs and fewer waves can penetrate away \citep{2008MNRAS.387..446T}.
For Rossby modes, there is no inner LR near the ICR which means the vanish of evanescent regions, so waves can penetrate readily in the planet.   
One can also identify the origins of two modes. The inertia-acoustic modess are excited within the BL and then propagate into the planet and the disk, whereas the Rossby modes are excited in the planetary atmosphere and then propagate outward. 

\begin{figure}[t]
    \centering
    \includegraphics[width=0.76\linewidth]{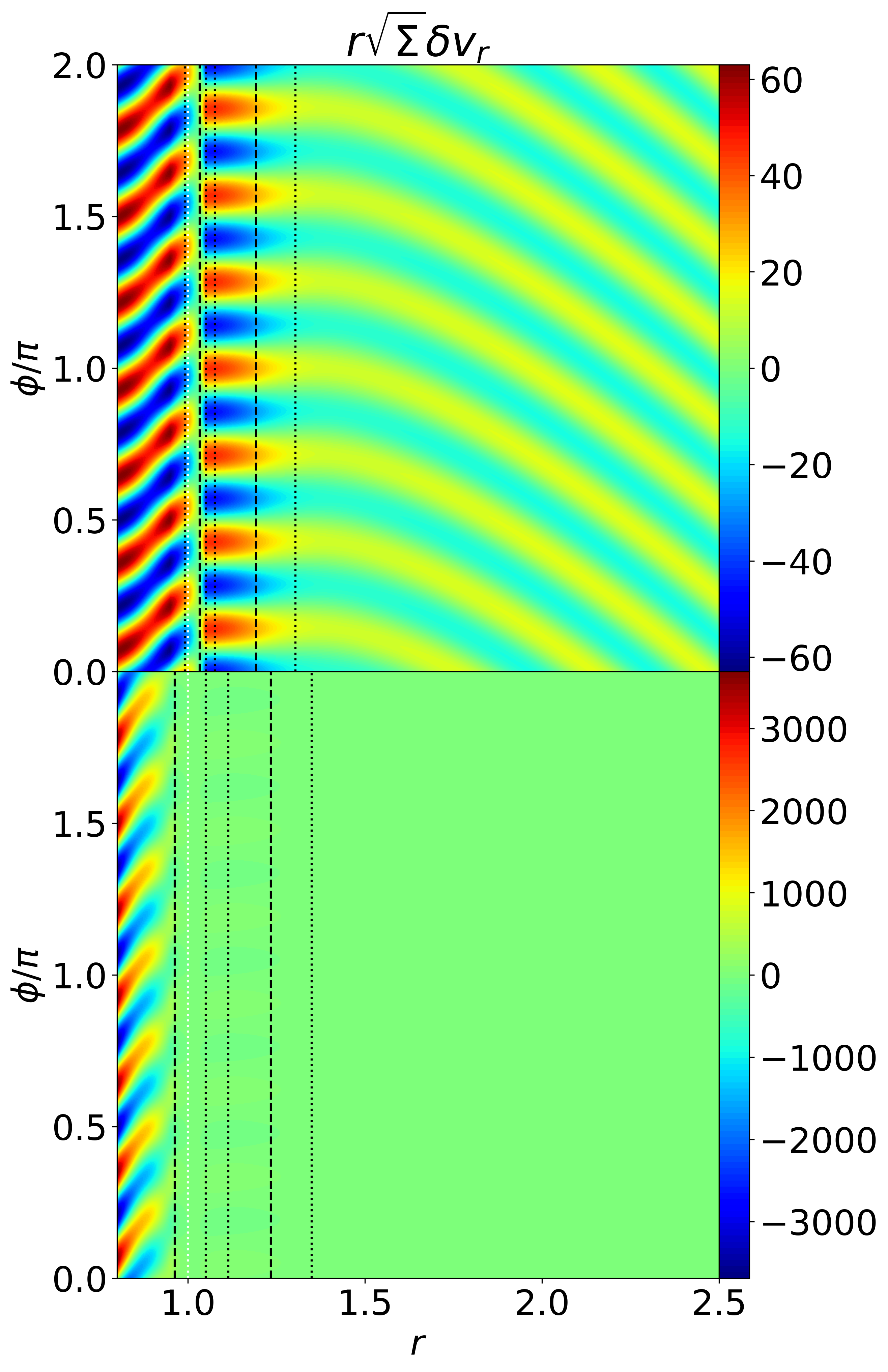}
    \caption{The 2D distribution of the physical variable, $r\sqrt{\Sigma} \delta v_r$. The upper panel shows the inertia-acoustic mode, which takes on inclined acoustic waves in the planet and density waves with trailing spiral arms in the disk. The lower panel is the Rossby mode, most of the energy is concentrated within the planet. The dashed lines indicate the corotation resonance, and the dotted lines indicate the Lindblad Resonance.}
    \label{energy_snapshot}
\end{figure}

In the upper panel of Figure \ref{energy_snapshot},  we show  the distribution of $r\sqrt{\Sigma}\delta v_r$ for inertia-acoustic modes which can be divided into three parts. 
Inside the planetary atmosphere with $r<1$, the propagating inertia-acoustic waves has non-zero $k_r$ and takes on inclined wave crests. 
However, between two CRs, there emerges some structures like bullets which turn out to be vortices in 2D distribution of pressure perturbation with the velocity field, as seen in Figure \ref{p-f_snapshot}.
Beyond the region, it is the density wave with trailing spiral arms as a result of $\Omega < \Omega_P$. 
Comparing our results to the numerical simulations by \cite{2022MNRAS.509..440C}, the pattern speed and the 2D wave spatial distribution are similar, implying the validity of our calculation.
We show the new Rossby mode for the spinning planets in the lower panel for comparison. The pattern is similar to the inertia-acoustic mode. But most of the energy is concentrated in the planet and vortices are less prominent.   

\subsection{Formation of Vortices}\label{Vortices}
\begin{figure}[t]
    \centering
    \includegraphics[width=0.75\linewidth]{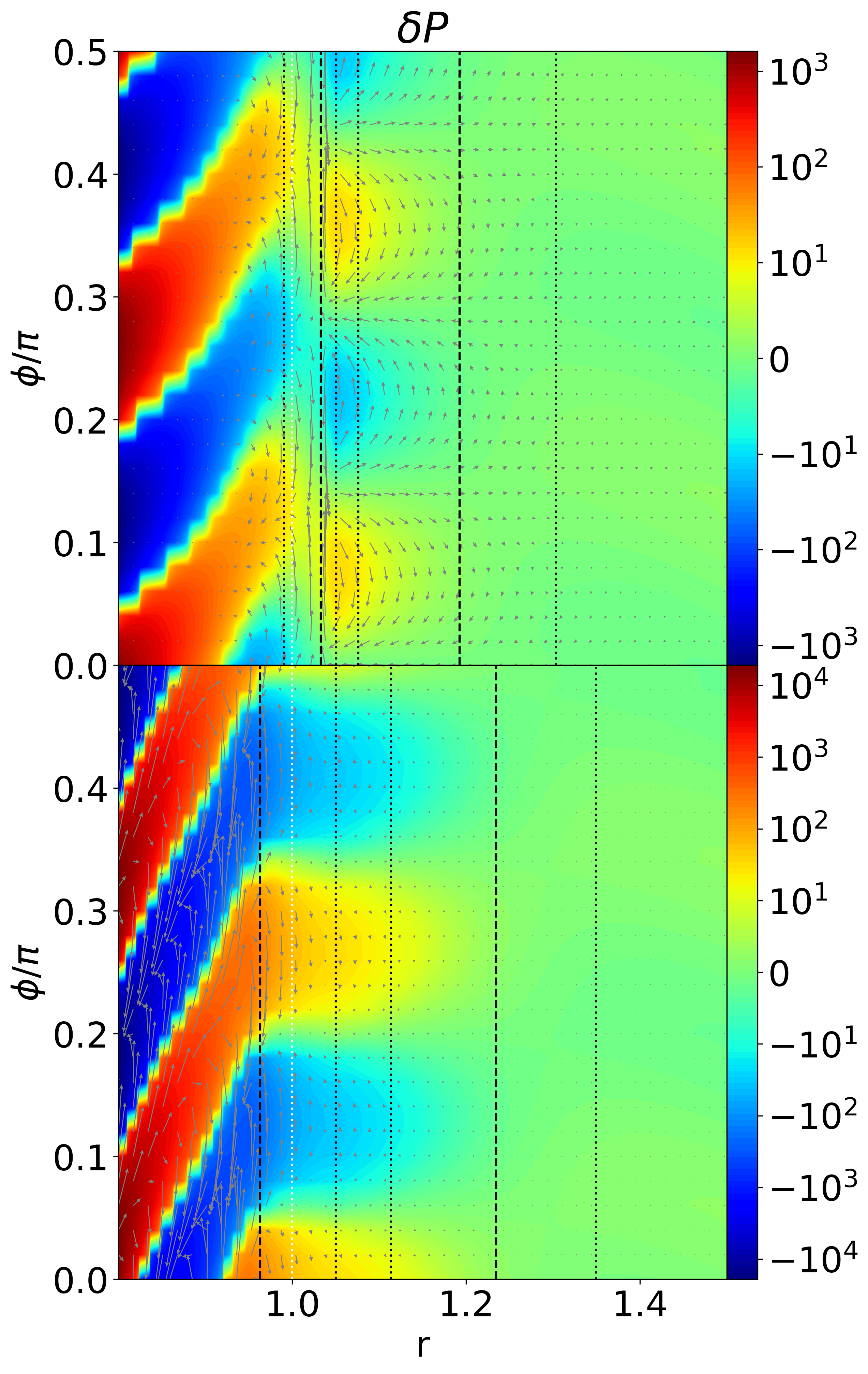}
    \caption{The 2D distribution of pressure perturbation together with the flow field indicated by arrows. The upper panel is the inertia-acoustic mode and the lower panel is the Rossby mode. The dashed lines indicate the corotation resonance, and the dotted lines indicate the Lindblad Resonance.
  }
    \label{p-f_snapshot}
\end{figure} 
\cite{2022MNRAS.509..440C,2022MNRAS.512.2945C} studied the spatial morphology of acoustic mode in detail and discovered some new vortex-driven modes in the vicinity of the BL. 
It is supposed that the formation of vortices within BL and those connecting spiral arms caused by density waves can be explained by the theory of Rossby wave instability (RWI). 
The generation of RWI requires an extremum in the radial profile of the key function 
\begin{equation}
    \mathscr{L}(r) \approx \frac{\Omega}{\kappa^2}\Sigma^{2/\Gamma-1}T^{2/\Gamma}
\end{equation}
\citep{1999ApJ...513..805L}. Here, $\Gamma$ is the adiabatic index and $T$ is the temperature.  
If we take into account the isothermal condition that $\Gamma=1$ and T remains constant, the criterion can be simplified further into $\mathscr{L}(r) \approx \Omega \Sigma / \kappa^2$. Retrospect to the Figure  \ref{surface_density}, we can know that the density profile is monotonous among the region, while the gradient of rotation speed has changed from zero to positive and then negative. 
It is clear that there is a minimum in the BL, which explains the emergence of vortices and spiral arms. 
Although the Papaloizou-Pringle instability can also result in the formation of vortices in compressible disks, which requires specific radial boundary conditions and angular momentum distribution, it does not hold for our model with a Keplerian disk\citep{1984MNRAS.208..721P}.

Figure \ref{p-f_snapshot} presents the distribution of pressure disturbances overlapped with the flow field described by $(\delta v_{r}, \delta v_{\phi})$. 
Between the ICR and the OCR, we can see that local pressure extrema lead to the formation of vortices and spiral arms.
For inertia-acoustic modes in the upper panel, materials between planetary surface and the ICR will directly fall into the planet.
However, the ICR impedes the exchange of substances inside and outside, which causes the formation of a strong shear field. 
This phenomenon is not observed for Rossby modes in the lower panel, where materials can travel freely through the ICR.

\subsection{The Transport of Mass and Angular Momentum }\label{Transport}
 To understand whether the wave can spin up the planet, an important indicator is the distribution of angular momentum flux (AMF). Generally, it can be decomposed into an advective term and a stress term. But for a wave, we just need to consider the latter term which can be expressed as 
\begin{equation}
    F_S(r)= 2\pi r \left< \Sigma v_r v_{\phi} \right> \ .
\end{equation}
Apart from the transport of angular momentum, the wave excited in the BL can also give rise to the accretion onto the planet. Thus, we measure the mass accretion rate by the definition  
\begin{equation}
    \dot M(r)=-2\pi r \left<\Sigma v_r \right> \ ,
\end{equation}
where $\left< ... \right>$ denotes the azimuthally averaged value.

In our linear analysis, the above equations can be written explicitly in terms of perturbation quantities \citep{2002ApJ...565.1257T,2008gady.book.....B}, namely
\begin{equation}
    F_S(r) = \pi r^2 \Sigma\left [ {\rm Re}( \delta v_r ) {\rm Re}(\delta v_{\phi} ) + {\rm Im}( \delta  v_r ) {\rm Im}( \delta v_{\phi} )  \right ] \ .
\end{equation}
Here, $F_S<0$ means inflow of angular momentum, while $F_S>0$ means outflow of angular momentum. Similarly, the accretion rate can be writtern as  
\begin{equation}
    \dot M (r) = - \pi r \left [ {\rm Re}( \delta \Sigma ) {\rm Re}(\delta v_r ) + {\rm Im}( \delta \Sigma ) {\rm Im}( \delta v_r )  \right ] \ ,
\end{equation}
 where $\dot M <0$ means decretion of mass, while $\dot M>0$ means accretion of mass.

\begin{figure}[t]
    \centering
    \subfigure[]
    {
    \centering
    \includegraphics[width=1\linewidth]{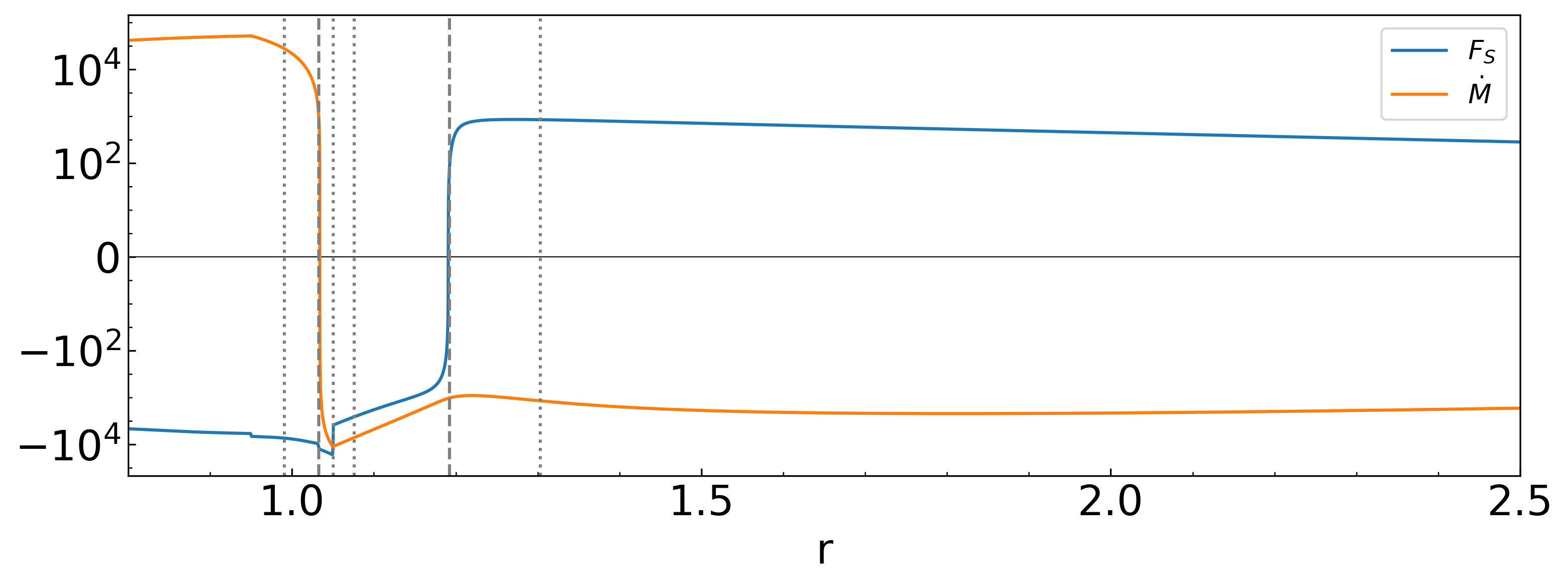}
    }
    \centering
    \subfigure[]
    {
    \centering
    \includegraphics[width=1\linewidth]{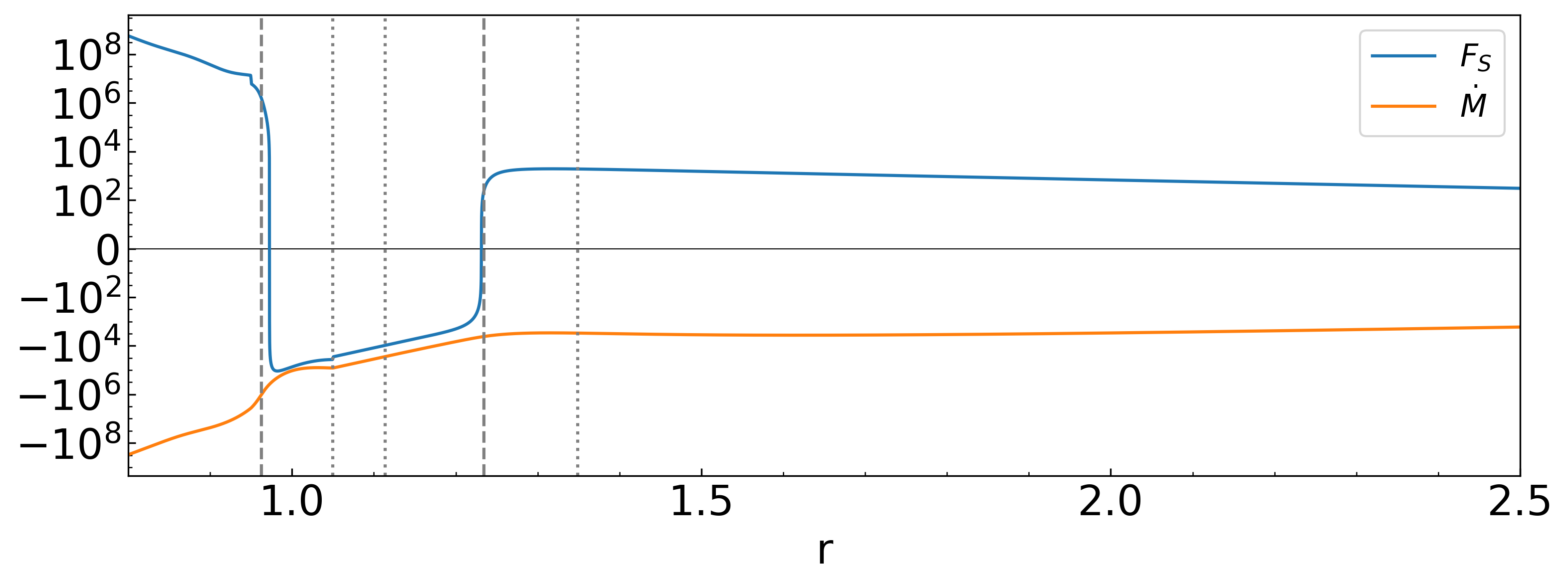}
    }
    \caption{Distributions of the angular momentum flux and the accretion rate, corresponding to inertia-acoustic modes in panel (a) and rossby modes in panel (b). The dashed lines indicate the corotation resonance, and the dotted lines indicate the Lindblad Resonance.}
    \label{AMF and Mdot}
\end{figure} 

The panel (a) and (b) in Figure \ref{AMF and Mdot} present the distribution of $F_S$ and $\dot{M}$ for inertia-acoustic modes and Rossby modes. 
Obviously, it can be divided into three regions according to the position of the ICR and the OCR.
Outside the OCR, both modes display positive $F_S$ and negative $\dot{M}$. 
This is because materials move slower than waves in the disk where $\Omega<\Omega_P$, the dissipation of waves can cause the angular momentum to be transported from the waves to the disk materials. For materials in a Keplerian disk, whose angular momentum is proportional to $r^{1/2}$, the acquired angular momentum will allow them to migrate outward. Hence, both modes lead to the expansion of the disk.   

Between the ICR and the OCR, both $F_S$ and $\dot{M}$ are negative for two modes. 
For inertia-acoustic modes, we can understand it in this way. As we mentioned before, both of the ICR and the OCR form a resonant cavity resulting from the presence of evanescent regions. It is like a potential well as described in quantum mechanics.
Majority of materials accumulate here and only a small fraction is able to penetrate into the planet, thereby creating local pressure centers outside the OCR, as shown in Figure \ref{p-f_snapshot}. Then, those pressure centers attract the surrounding material to evolve into vortices, which induce more materials to migrate inward.
It is worth noting that despite $\dot{M}$ outside the ICR is consistently negative, its minimum occurs within the BL. Thus, a gap will form in the density profile and this phenomenon has been confirmed by numerical simulations. 
While for Rossby modes, the obstruction of the ICR to waves gets weaker because of the disappearance of evanescent regions. Therefore, vortices are less pronounced than inertia-acoustic modes, as shown in Figure \ref{p-f_snapshot}.

The main difference between these two modes is that they manifest opposite properties inside the ICR.
$F_S$ is negative and $\dot{M}$ is positive for inertia-acoustic modes, so waves will give rise to accretion of gas and planetary spin-up. On the other hand, $F_S$ is positive and $\dot{M}$ is negative for Rossby modes, so waves will give rise to decretion of gas and planetary spin-down. It is the vastly different nature of these two modes in the planet that leads to the generation of terminal rates.

\subsection{Growth Rate Dependence on Various Parameters}\label{Impact}

\begin{figure}[t] 
    \centering
    \subfigure[]
    {
    \centering
    \includegraphics[width=0.8\linewidth]{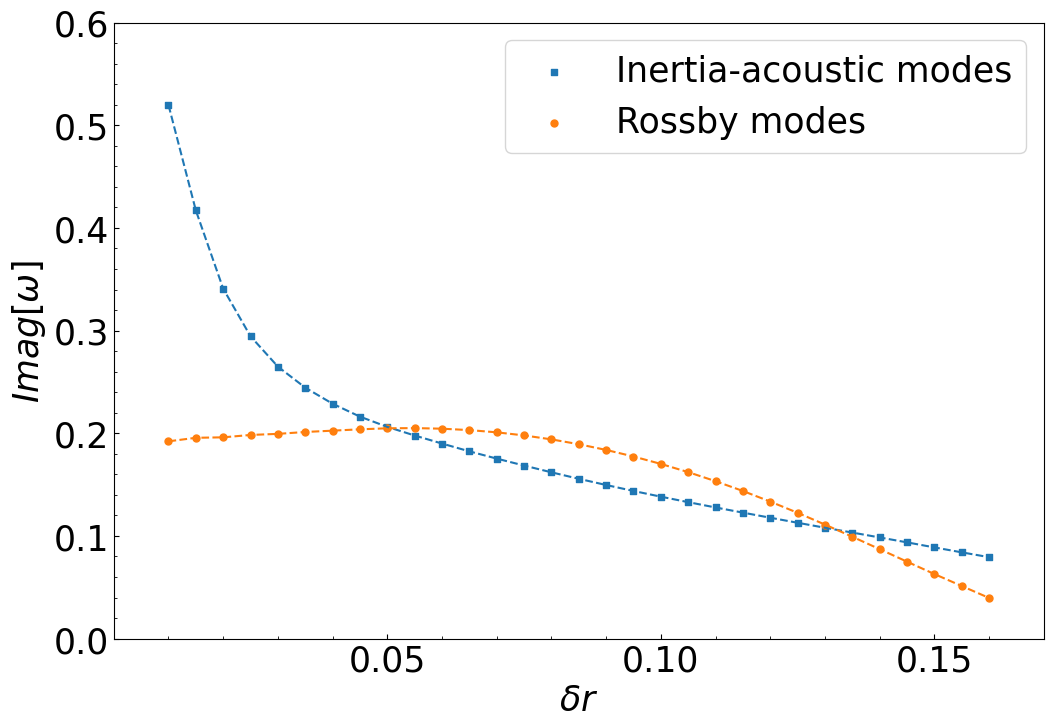}
    }
    \centering
    \subfigure[]
    {
    \centering
    \includegraphics[width=0.8\linewidth]{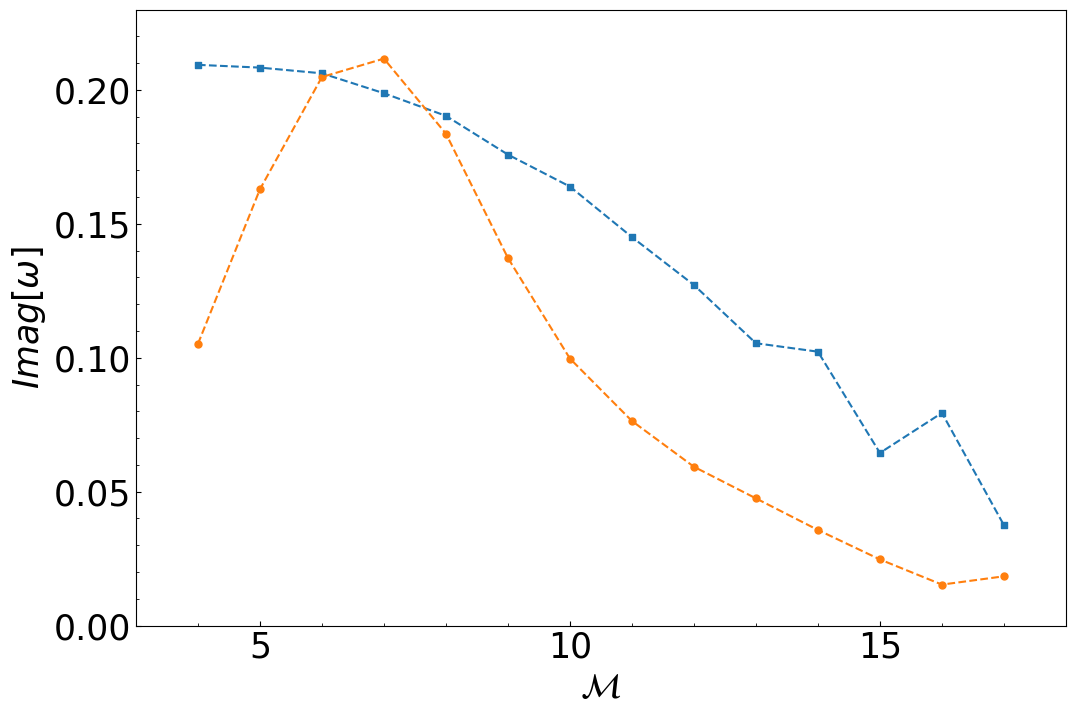}
    }    
\subfigure[]
    {
    \centering
    \includegraphics[width=0.8\linewidth]{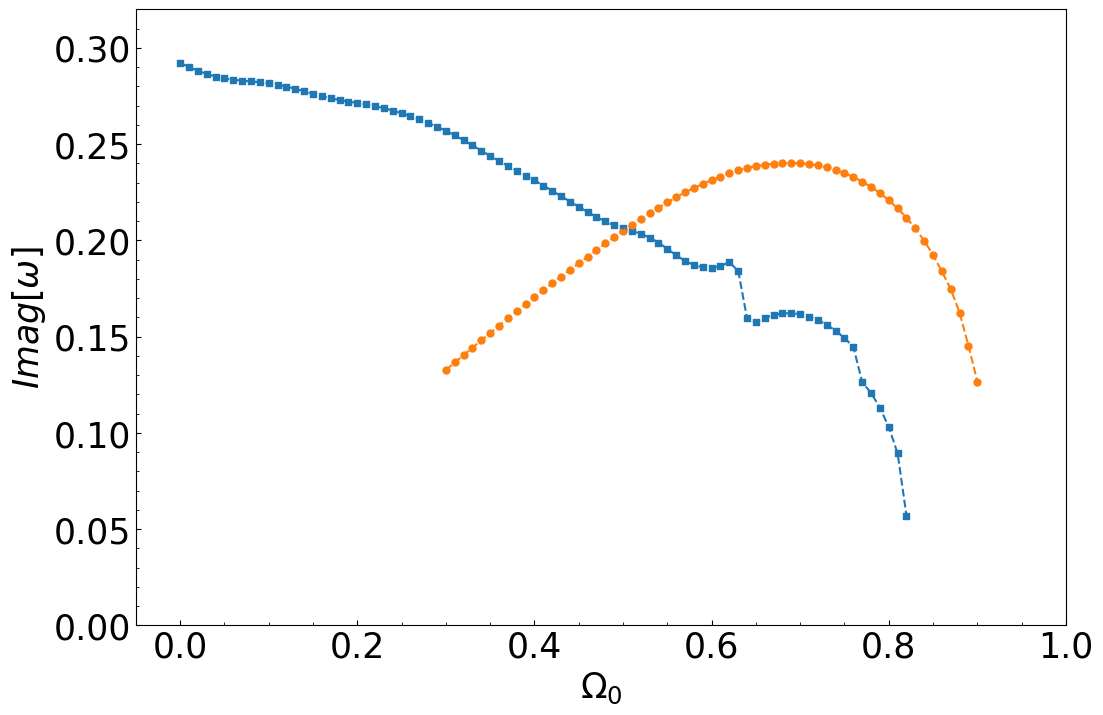}
    }
\caption{Dependence of the growth rate Im$[\omega]$ of inertia-acoustic modes and Rossby modes on the width of BL $\delta r$, the Mach number $\mathcal{M}$ and the spin rates $\Omega_0$. The results of inertia-acoustic modes are represented in blue dots, while the results of Rossby modes are represented in orange dots.}
\label{inc}
\end{figure}

Based on the behaviors of AMF and accretion rate of two distinct modes, we know that the two modes play opposite roles for the terminal spins. Thus, the relative strength of the two modes would determine the spin-up or spin-down of planets. In this section, we will explore the dependence of the growth rate on the relevant parameters.

To compare the strength of inertia-acoustic modes and Rossby modes, we choose a particular mode with $m=25$, with $\mathcal{M}=6, \delta r=0.05$ and $\Omega_0=0.5$ as our fiducial parameters. In the subsequent exploration of the parameter space, we change only one of them. 
Figure \ref{inc} shows the dependence of the growth rate Im$[\omega]$ on the relevant parameters: panel (a) for the width of BL $\delta r \sim 0.1 - 0.16$; panel (b) for the Mach number $\mathcal{M} \sim 4 - 17  $ and panel (c) for the spin rates $\Omega_0 \sim 0 - 0.9 $. The results of inertia-acoustic modes are represented in blue dots, while the results of Rossby modes are represented in orange dots.

As can be seen from panel (a), there is a clear trend of the growth rate decrease with width of BL for inertia-acoustic modes, Im$[\omega] \propto \delta r^{-1}$. 
While for Rossby modes, the growth rate peaks at around $\delta r =0.05$ and then drops sharply with increasing $\delta r$.
Comparing the two modes, it is interesting to note that the growth rate of Rossby modes exceeds that of inertia-acoustic modes at $\delta r \sim 0.05 - 0.13$.  This suggests that although inertia-acoustic modes are stronger than Rossby modes, it will be dominated by Rossby modes for certain BL width. 

A similar trend appears in panel (b), but the growth rate of inertia-acoustic modes decreases more linearly.
The growth rate of Rossby modes continues to increase before $\mathcal{M}=7$ and then decreases sharply with increasing $\mathcal{M}$. 
 A comparison of the two results also shows that the growth rate of Rossby modes exceeds that of inertia-acoustic modes at around $\mathcal{M}-6\sim 8$.   
 In general, inertia-acoustic modes are stronger than Rossby modes, but Rossby modes can dominate the BL in certain Mach number ranges.

A more interesting observation is about the dependence of growth rates on the spin rates $\Omega_0$ in panel (c).
The growth rate of the inertia-acoustic mode decreases with increasing $\Omega_0$ until it disappears at $\Omega_0=0.83$. Rossby modes start to emerge only at a certain spin rate, peaking at $\Omega_0=0.7$.
When $\Omega_0 <0.5$, the inertia-acoustic modes dominate. When $\Omega_0 >0.5$, the Rossby modes dominate. The terminal spin may be determined by the competition between the two modes. 
Generally speaking, inertia-acoustic modes are dominating at low $\Omega_0$ while Rossby modes are more prominent at high $\Omega_0$.

In real cases, the system is a combination of multiple modes, so the above analysis is insufficient to give an accurate terminal rate. However, these results suggest a strong correlation between the dominant mode and the spin rate, although the width of BL or the Mach number can also play a significant role to some extent.

\section{Discussion}\label{Discussion}
Results of previous sections compare inertia-acoustic modes with Rossby modes on morphology, angular momentum transport, accretion rate and growth rate. We now explain the solution of global modes and discuss the connection between theory and numerical simulations.
\subsection{Interpretation of Solutions} 
\begin{figure}[t]
    \centering
    \subfigure[]
    {
    \centering
    \includegraphics[width=0.9\linewidth]{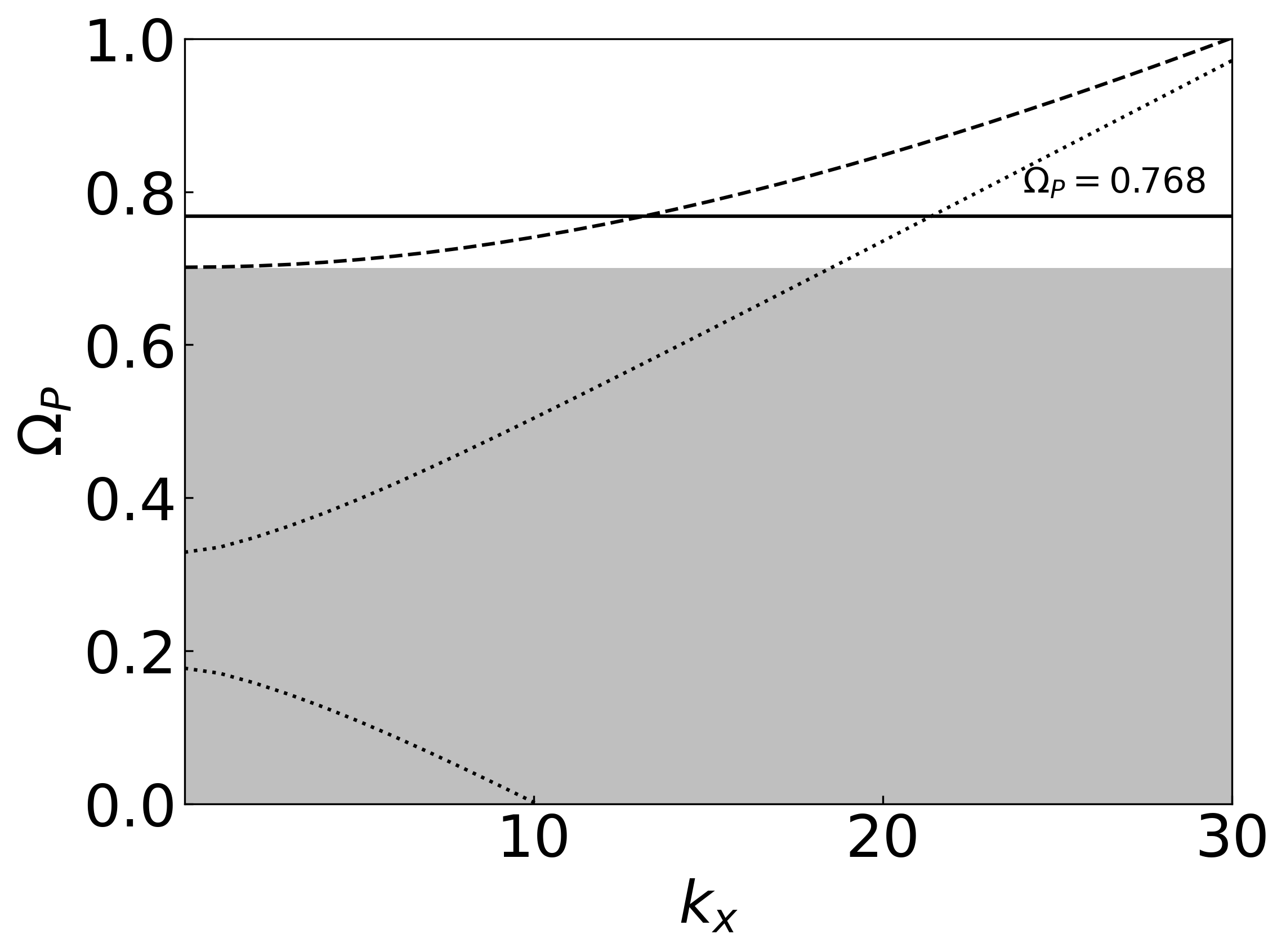}
    }
    \subfigure[]
    {    
    \centering
    \includegraphics[width=0.9\linewidth]{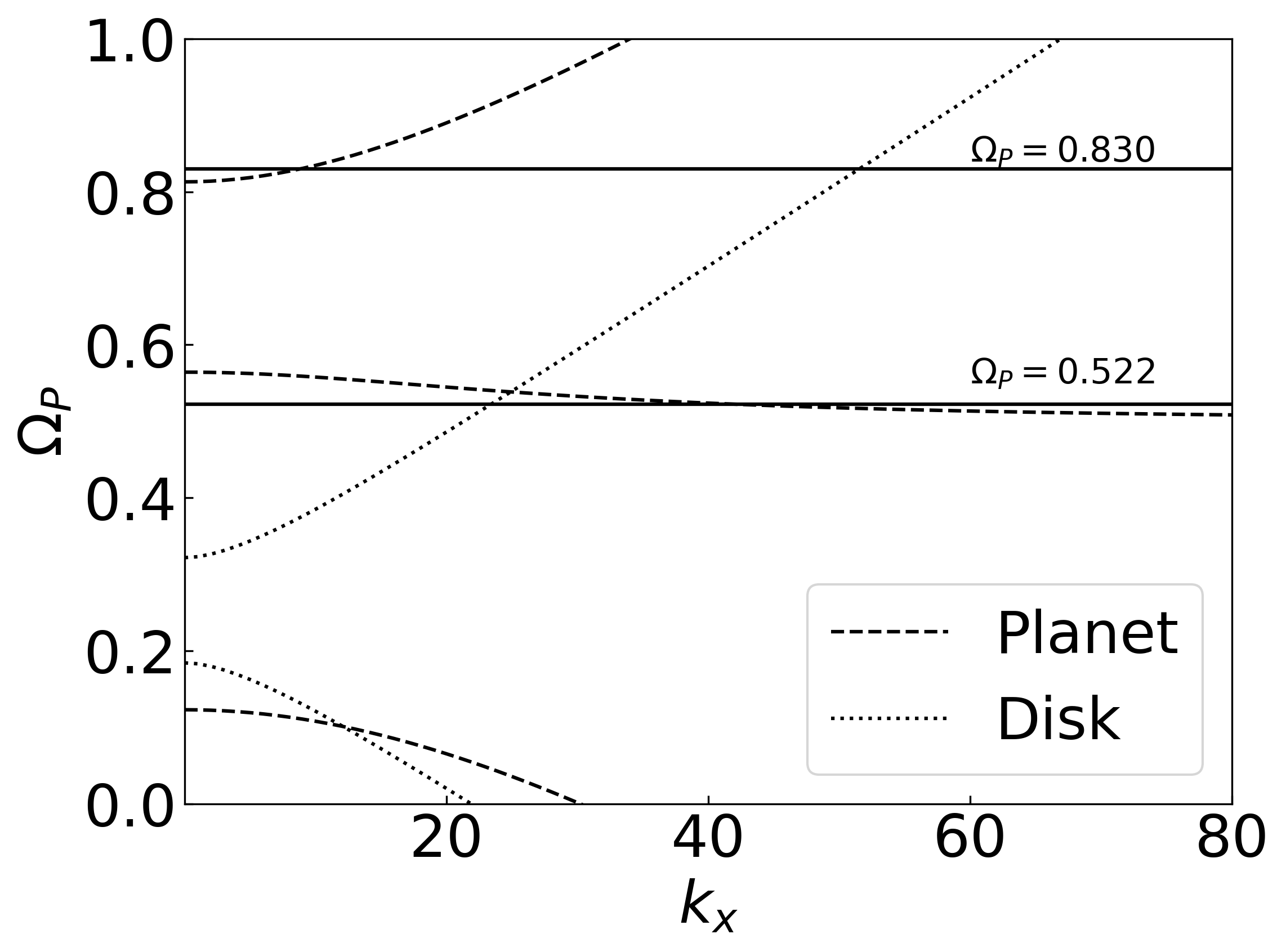}
    }    
\caption{Diagrams of $\Omega_P - k_x$. Dashed lines represent the dispersion relation of the planetary atmosphere, while dotted lines are the dispersion relation of the disk. The black lines are $\Omega_P$  calculated by our program under parameters (a) $m=7, \delta r = 0.05, \mathcal{M} = 6$ and $\Omega_0=0$. (b) $m=15, \delta r = 0.05, \mathcal{M} = 6$ and $ \Omega_0=0.5$. Solutions in the shaded zone are forbidden.}
\label{Omega_P-k_x}
\end{figure}

Although global modes cannot be solved by local analysis, the numerical solutions can be understood from the local dispersion relation.
Inspection of the equation \eqref{inner dispersion} shows that if $\Omega=0$ only acoustic modes appear. If $\Omega\neq0$ it has three different solutions. Two correspond to the inertia-acoustic mode and one corresponds to the Rossby mode. When it comes to the equation \eqref{Outer dispersion}, it keeps the form of the density wave regardless the planet spins or not. 
A global mode should satisfy the above two dispersion relations simultaneously, so we put them on the same graph, as shown in Figure  \ref{Omega_P-k_x}. 
According to the parameters in Table \ref{Results}, we plot the dispersion relations for a non-rotating planet in panel (a) and a spinning planet with $\Omega_0=0.5$ in panel (b).
The dashed lines represent the dispersion relation of the planetary atmosphere, and the dotted lines are the dispersion relation of the disk. 
We also plot the pattern speed calculated by our program in black lines.

For a fixed $\Omega_P$, only the values that intersect with both inner and outer dispersion curves are valid. Thus, solutions in the shaded zone are forbidden, as shown in panel (a). This is verified by numerical results.
Besides, the $k_1$ and $k_2$ acquired from the intersection of Figure \ref{Omega_P-k_x} are in good accordance with the results of Table \ref{Results}. 
This means that global modes are essentially combinations of local waves, though they cannot be solved by local analysis.

\subsection{Super-reflection and Resonant Cavity}
It has been observed that vortices and spiral arms will form at the beginning \citep{2022MNRAS.509..440C,2022MNRAS.512.2945C}, as we proved in $\S$\ref{Vortices}. However, we noted that such phenomena will disappear for a while as the evolution of systems and reoccur at late times. Combining theoretical and numerical simulation results, our analysis is as follows.

Due to the sharp shear layer in the initial rotation profile, global modes can emerge during the linear phase, which include the inclined waves in the planetary atmosphere and the formation of vortices and spiral arms in the disk. The angular momentum carried by waves migrates inward and causes the planet to spin faster. At the same time, accretion of surrounding materials leads to the formation of gaps around the planet. Next, the growth rate of global modes will gradually weaken resulting from the expansion of BL over time. Meanwhile, majority of waves are trapped in a resonant cavity due to the super-reflection of two CRs \citep{2008MNRAS.387..446T}. Those outgoing waves reflect at the ILR and then form a criss-cross structure with the original waves. Such a phenomenon named as lower mode evinces the beginning of nonlinear phase. Furthermore, deeper and deeper gaps excite the Rossby wave instability again, so vortices and spiral arms emerge again. 

\subsection{Spin-down and Decretion}
Outflow of angular momentum and decretion of gas at high $\Omega_0$ have been found in \cite{2021ApJ...921...54D}. At $\Omega_0=0.5$ and $\Omega_0=0.7$, they found radial inflow for $r<2$ and radial outflow for $r>2$ at the mid-plane in the inner disk region. However, at $\Omega_0=0.8$, they found radial outflow at the mid-plane for all radii. Such phenomena can be attributed to the transition of dominating modes from inertia-acoustic modes to Rossby modes. It is noting that we still adopt the initial conditions with maximum in $\Omega(r)$, varying from their initial conditions that $\Omega(r)$ should be monotonically decreasing. This difference is likely to be a consequence of introduction of viscosity, and they are carrying out 2D simulations in axisymmetric geometry.

\subsection{Coexistence of inertia-acoustic and Rossby Modes}
In $\S$\ref{Impact}, we have pointed out that the inertia-acoustic mode are dominating among all modes at low $\Omega_0$, so planet spins up accompanying by accretion of gas. However, at high $\Omega_0$, Rossby modes complicates the problem, since they can give rise to the planetary spin-down  and gas decretion. The two modes always exist together and it is hard to distinguish them separately, especially in numerical simulations. Nonetheless, as a whole, the planet will spin up and accrete gas provided that most of modes are dominated by inertia-acoustic modes. Otherwise, the planet will spin down and decrete gas provided that most of modes are dominated by Rossby modes. A system can even switch between these two states when we take into account the effect of $\delta r$ and $\mathcal{M}$. 

One plausible example can be found in \cite{2021MNRAS.508.1842D}, who studied into the terminal spin of planet and found that the efficiency of angular momentum transport and the accretion rate will decrease largely at high $\Omega_0$. In his numerical simulations at $\Omega_0 = \{0.4, 0.8 \}$ when $\mathcal{M}=6$, where the author plotted the temporal evolution of angular momentum current $C_S$. What he did not notice is that $C_S$ becomes positive near $r=1$ at some early stages. However, such phenomena disappear when $\mathcal{M}=10$. 
By our theory, we can explain it this way. At the very beginning, $\delta r$ is not very large so inertia-acoustic modes are dominating among all modes, so $C_S$ is negative inside planet. 
After a period of time, $\delta r$ expands to a certain extent that is enough for Rossby modes to dominate among all modes. Thus, $C_S$ is positive near planetary surface. 
At late times, further expansion of $\delta r$ weaken the Rossby modes, and inertia-acoustic modes dominate among all modes, with negative $C_S$ inside planet again. 
On the other hand, inertia-acoustic modes are more prominent at high $\mathcal{M}$, so $C_S$ is almost always negative inside planet when $\mathcal{M}=10$.

\section{Summary}\label{Summary}
We perform a linear analysis of isothermal planet-disk systems with the BL. Using the WKB approximation, we derive the dispersion relations and corresponding group velocities. 
The dispersion relation of the disk is a quadratic equation that is in agreement with the theory of density waves, while the dispersion relation of the planet is a cubic equation if we take into account rotation. 
A global mode should satisfy the above two relations simultaneously, so we concluded that there should be three types of global modes when the central body is rotating. 
Based on dispersion relations and group velocities, two of them are identified as the inertia-acoustic mode with opposite directions of propagation, but another one with low frequency respect to the rotating planet is referred to as the Rossby mode.

The spatial morphology of two modes is similar, including the pattern of inclined waves in the planetary atmosphere, vortices near the BL and spiral arms in the disk. 
For each mode with pattern speed $\Omega_P>\Omega_0$, there are always two corotation resonances, the ICR and the OCR. 
When both of the CRs locate in the disk, a resonant cavity can form due to the presence of evanescent regions and super-reflection of the CR.
Thus, waves are excited within the BL and propagate away the BL for inertia-acoustic modes.
However, the evanescent region vanishes when the ICR locates in the planet, so waves can propagate freely from the planet to the disk for Rossby modes. 

Next, we study the AMF $F_S$ and accretion rate $\dot M$ caused by waves. Both of them shows positive $F_S$ and negative $\dot M$ outside the OCR, implying the expansion of disks. Between two CRs, negative $F_S$ and the minimum of negative $\dot M$ within the BL mean the formation of gap around the planet. The main differences occur inside the ICR including the planetary atmosphere. 
\begin{itemize}
\item{$F_S$ is negative and $\dot M$ is positive for inertia-acoustic modes, so waves will spin up planets and lead to gas accretion.}
\item{$F_S$ is positive and $\dot M$ is negative for Rossby modes, so waves will spin down planets and lead to gas decretion.
}
\end{itemize}
We also study the growth rate of two modes as a function of the width of BL $\delta r$, the Mach number $\mathcal{M}$ and the spin rate $\Omega_0$. Our key results can be summarized as follows.

\begin{itemize}
    \item{For various $\delta r$ or $\mathcal{M}$, inertia-acoustic modes are stronger than the Rossby modes, but Rossby modes can be prominent at certain ranges.}
\item{For various $\Omega_0$, inertia-acoustic modes are dominating at low $\Omega_0$ while Rossby modes are more prominent at high $\Omega_0$.
}
\end{itemize}
Whether the planet spins up or spins down depends on the dominate mode. 
The turning point from inertia-acoustic modes dominant to Rossby modes dominant can be considered as the terminal spin.
However, it is hard to determine the exact value of terminal spins since our analysis is limited to a single mode.  
Non-linear features such as weak shocks \citep{2012ApJ...752..115B} or merging vortices \citep{2022MNRAS.509..440C,2022MNRAS.512.2945C}, their impacts on the local angular momentum transport and accretion rate are also beyond the scope of linear analysis. Further study on the evolution of two modes by numerical simulations would contribute to our understanding of their respective roles on limiting the spin rate.
In addition, inclusion of other factors like magnetic fields or cooling effects could also make the problem more intriguing, and these would be discussed in our subsequent papers \citep{2009ApJ...702...75Y,2022MNRAS.514.1733H}.

Our theory reveals the underlying dynamic mechanism of terminal spins, deepening our understanding of the angular momentum transport and the accretion process in planet-disk systems with the BL.

\section*{Acknowledgement}
We thank the anonymous referee for helpful suggestions that have greatly improved this paper. This work has been supported by the National SKA Program of China (Grant No. 2022SKA 0120101) and the National Key R$\&$D Program of China (No. 2020YFC2201200), the science research grants from the China Manned Space Project (No. CMSCSST-2021-B09 and CMS-CSST-2021-A10), and opening fund of State Key Laboratory of Lunar and Planetary Sciences (Macau University of Science and Technology) (Macau FDCT Grant No. SKL-LPS(MUST)-2021-2023). C.Y. has been supported by the National Natural Science Foundation of China (Grant Nos. 11521303, 11733010, and 11873103).

\bibliography{sample631}{}
\bibliographystyle{aasjournal}



\end{document}